\documentclass[aps,twocolumn,showpacs,prd]{revtex4}
\usepackage{graphicx}
\usepackage{amssymb}
\usepackage{amsmath}

\newcommand{\be}{\begin{equation}}
\newcommand{\ee}{\end{equation}}
\newcommand{\ben}{\begin{eqnarray}}
\newcommand{\een}{\end{eqnarray}}
\newcommand{\bes}{\begin{subequations}}
\newcommand{\ees}{\end{subequations}}

\input{tcilatex}
\begin{document}

\title{Generalized self-dual Chern-Simons vortices}
\author{D. Bazeia$^{a}$, E. da Hora$^{a}$, C. dos Santos$^{b,c}$, and R.
Menezes$^{b,d}$}
\affiliation{$^{a}$Departamento de F\'{\i}sica, Universidade Federal da Para\'{\i}ba,
58051-970 Jo\~{a}o Pessoa, Para\'{\i}ba, Brazil}
\affiliation{$^{b}$Centro de F\'{\i}sica e Departamento de F\'{\i}sica, Faculdade de Ci%
\^{e}ncias da Universidade do Porto, 4169-007 Porto, Portugal}
\affiliation{$^{c}$Departamento de F\'{\i}sica, Universidade de Santiago de Compostela,
15782 Santiago de Compostela, Spain}
\affiliation{$^{d}$Departamento de Ci\^encias Exatas, Universidade Federal da Para\'{\i}%
ba, 58297-000 Rio Tinto, Para\'{\i}ba, Brazil}

\begin{abstract}
We search for vortices in a generalized Abelian Chern-Simons model with a
non-standard kinetic term. We illustrate our results plotting and comparing
several features of the vortex solution of the generalized model with those
of the vortex solution found in the standard Chern-Simons model.
\end{abstract}

\maketitle


\section{Introduction}

The study of vortices in planar Chern-Simons (CS) models has been pioneered
in Refs.~\cite{csv1,csv2,csv3}. Since then, a lot of investigations on
Chern-Simons vortices have been done; see, e.g., \cite%
{dunne1,dunne2,sakai,schap}. During the past years, however, theories with
non-canonical kinetic term, named generalized of $k$-field models, have been
intensively studied. Generically, their applications have been found in
strong interaction physics, with the Skyrme \cite{Skyrme1} and Skyrme-like
models \cite{Skyrme2,Skyrme3,Skyrme4,Skyrme5,bmm,Skyrme6}, and also in
cosmology with the so-called $k$-essence models \cite{kessence1,kessence2}. $%
k$ fields change the way the fields approach their vacuum values, allowing
thereby, for instance, the existence of solitons which approach their vacuum
values in a power-like instead of an exponential fashion and which therefore
have a compact support \cite{Adam1,Adam2,Adam3,Adam4,Adam5}. Also, $k$%
-theories allow to avoid Derrick's theorem \cite{Derrick} increasing the
chances to find soliton solutions in symmetry-breaking models. By this way,
several $k$-topological defects were already studied by several authors \cite%
{Babichev1,Jin,Sarangi,bmp,bglm,Bazeia0,Bazeia3,Babichev2,Babichev3} and the
overall conclusion is that their properties can be quite different from the
standard ones depending specifically of the choice made for the kinetic term.

The non-linear effects in $k$-theories make the equations of motion more
difficult to solve and therefore we will focus only on BPS vortex solutions
which minimize the energy. They can be found by minimizing the energy
functional \cite{Bogomol'nyi1,Prasad1} or equivalently by using the
conservation law for the energy-momentum tensor combined with the boundary
conditions that require finite energy for the vortex \cite{Vega1}. This
method combined with supersymmetry arguments allows for the first order
formalism developed in ref. \cite{Bazeia0}, used to obtain BPS global $k$
defects in one dimension and whose linear stability is proved analytically.
In particular, when considering perturbative corrections to the canonical
kinetic term, the authors found linearly stable solitons which, as expected,
do not differ physically from the standard kinks, as they have the same
energy, even though their width and energy densities are different. They
also found kink solutions through a specific combination for the
non-canonical kinetic term and the potential. Finally, the method was used
to obtain topological compactons \cite{C1,C2,C3,C4}, e.g., solitons which
approach the vacuum values at finite distance, confirming the results of
ref.~\cite{Adam2}. Thus, an important conclusion of this work is that even
being a first order formalism, it is suitable for the study of
nonlinearities in the kinetic term. More recently, supersymmetric extensions
of $k$-field models has been introduced \cite{bmp}.

The most important aim of the present work is to generalize to vortices the
first order formalism developed in ref.~\cite{Bazeia0}. We take a $(2,1)$
Abelian Chern-Simons model with a non-canonical kinetic term for the complex
scalar field. We apply the method of ref.~\cite{Vega1} to obtain the first
order equations of motion, and then search for vortex solutions choosing the
usual rotationally symmetric \textit{Ansatz} for the scalar and gauge
fields. First, we check that all vortex solutions of the Bogomol'nyi
equations, i.e., BPS vortices, are physical requiring that they minimize the
action. We then study analytically the BPS vortex equations and present a
vortex solution for a specific choice for the non-canonical kinetic term.
Finally, we compare our results with the BPS vortex solutions obtained in
refs.\cite{csv1,csv2}. We use standard conventions, taking a $(2,1)$
spacetime with a plus minus signature for the Minkowski metric $\left(
+--\right)$ and using bold style for the spatial components of 3-vectors.


\section{The model}

We take an extension of the vortex model suggested by refs.\cite{csv1,csv2},
which has the standard form 
\begin{equation}
\mathcal{L}_{S}=\frac{k}{4}\epsilon ^{\alpha \beta \gamma }A_{\alpha
}F_{\beta \gamma }+\left\vert
D_{\mu}\varphi\right\vert^{2}-V\left(\left\vert \varphi \right\vert \right),
\end{equation}
where $k$ is a constant, $\varphi$ is the complex Higgs field and $V\left(
\left\vert \varphi \right\vert \right)$ is its potential. Also, $%
F_{\alpha\beta}=\partial_\alpha A_\beta-\partial_\beta A_\alpha$ and $D_{\mu
}=\partial _{\mu }+ieA_{\mu }$, with $e$ being the electric charge. Here we
are using $A^{\beta }=\left( A^{0},\mathbf{A}\right)$, and the electric and
magnetic fields are given by $\mathbf{E^i=}F^{i 0}=-\dot{{\mathbf{A}}^i}%
-\nabla_i A^{0}$ and $\mathbf{B}=\overrightarrow{\nabla }\times\mathbf{A}$,
respectively.

We modify this model by changing the canonical kinetic term of the scalar
field, as described by the new Lagrangian density 
\begin{equation}
\mathcal{L}_{G}=\frac{k}{4}\epsilon^{\alpha \beta \gamma}A_{\alpha }F_{\beta
\gamma }+w(|\varphi|)\left\vert D_{\mu }\varphi \right\vert
^{2}-V\left(\left\vert \varphi\right\vert \right),  \label{l2}
\end{equation}
where $w(|\varphi|)$ is, in principle, an arbitrary function of the complex
scalar field. Note that the non-canonical term in the Lagrangian density, in
the limit $w(|\varphi|)\to1$, leads us back to the standard Chern-Simons
model.

It is convenient for the study of vortices to write all the variables in
dimensionless units. For that we take $x^\mu\rightarrow x^\mu/M$, where $M$
is a mass scale of the model. Also, we take the two parameters $k$ and $e$
as: $k\rightarrow M k$ and the electric charge $e\rightarrow M^\frac12 e$.
In this case, we get 
\begin{equation}
\varphi\rightarrow M^{\frac12}\varphi\;\; A_{\gamma } \rightarrow M^\frac12
A_{\gamma },
\end{equation}
and we can write $\mathcal{L}_G\rightarrow M^3\mathcal{L}_g$, with $\mathcal{%
L}_g$ being the Lagrangian density to be used from now on.


\section{Equations of motion}

The equations of motion for the gauge fields are given by 
\begin{equation}
\frac{k}{2}\epsilon ^{\mu \beta \gamma }F_{\beta \gamma }=-J^{\mu }
\label{maria1}
\end{equation}
where $J^{\mu }=(\rho ,\mathbf{j})$ is the current density given by 
\begin{equation}
J^{\mu }=i\,e\,w\,\left[ \varphi \left( D^{\mu }\varphi \right) ^{\ast
}-\varphi ^{\ast }D^{\mu }\varphi \right].
\end{equation}

The time and spatial components of eq.~(\ref{maria1}) are, for static field
configurations, 
\begin{equation}
kB=\rho=2e^{2}A_{0}\left\vert \varphi \right\vert ^{2}w(\left\vert \varphi
\right\vert)  \label{csgl1}
\end{equation}%
and 
\begin{equation}
k\mathbf{E}^a=\epsilon_{b a}\,j^b ,  \label{gl0}
\end{equation}%
which show that the electric charge density is proportional to the magnetic
field while the density current is perpendicular to the electric field. This
fact is important for the phenomenological applications of Chern-Simons
theories as effective field theories for the quantum Hall effect \cite{Hall1}%
.

The equation of motion for the scalar field is given by 
\begin{equation}
w\,\Box \varphi +\partial_\mu w D^{\mu }\varphi +\frac{\partial V} {\partial
\varphi ^{\ast }}-\left\vert D_{\mu }\varphi \right\vert ^{2}\,\frac{%
\partial\,w} {\partial \varphi ^{\ast }} =0  \label{sf3}
\end{equation}
with 
\begin{equation}
\Box \, \varphi = \frac{1}{\sqrt {g}}D^\mu \left[\sqrt{g}\,D_\mu\,\varphi%
\right]
\end{equation}
where $g$ is the determinant of the metric.

These second order differential equations will be reduced to first order
ones by using the method developed in \cite{Vega1}. For that, we need to
obtain the components of the energy-momentum tensor, given by 
\begin{equation}
T_{\lambda\rho}=\frac{2}{\sqrt{g}}\frac{\partial \left[ \sqrt{g}\mathcal{L}%
_g^{\prime}\right]}{\partial g^{\lambda \rho }},
\end{equation}
where $\mathcal{L}_g^{\prime}$ excludes the Chern-Simons term of the
Lagrangian density, since it does not contribute to the energy-momentum
tensor. For the vortex they become 
\begin{equation}
T_{\lambda \rho }=-g_{\lambda \rho }\mathcal{L}_g^{\prime}+2w\Psi_{\lambda
\rho }  \label{cset1}
\end{equation}%
with 
\begin{equation}
\Psi_{\lambda \rho }=\frac{1}{2}D_{\mu}\varphi
\left(D_{\nu}\varphi\right)^{\ast}\left[\delta_{\lambda\mu}
\delta_{\rho\nu}+\delta_{\lambda\nu}\delta_{\rho\mu}\right]
\end{equation}

Writing explicitly the components of the energy-momentum tensor one obtains 
\begin{eqnarray}
\varepsilon\equiv T_{00}&=& 2e^{2}w A_{0}^{2}|\varphi|^{2}-w
|D_{\mu}\varphi|^{2}+V  \label{csed2} \\
\mathcal{P}_{1}\equiv T_{11}&=& w\,|D_{\mu}\varphi|^{2}+2w
|D_{1}\varphi|^{2}-V \\
\mathcal{P}_{2}\equiv T_{22} &=& w\,|D_{\mu}\varphi|^{2}+2w
|D_{2}\varphi|^{2}-V \\
T_{01}&=&A_{0} j_1 \\
T_{02}&=&A_{0} j_2 \\
T_{12}&=& w\,\left[(D_{1}\varphi)(D_{2}\varphi)^{\ast}\!+\!(D_{1}\varphi)^{%
\ast}(D_{2}\varphi)\right]
\end{eqnarray}
where 
\begin{eqnarray}
|D_{i}\varphi|^{2}&=&|(\partial_{i}+ieA_{i})\varphi|^{2};\;\;\; i=1,2.
\end{eqnarray}
Now setting the vortex stability condition (see \cite{Vega1}) 
\begin{equation}
\mathcal{P}_{1}=\mathcal{P}_{2}=0  \label{p1}
\end{equation}
we get to the first order equations of motion 
\begin{eqnarray}
D_{\pm }\varphi &=&0  \label{cesb1} \\
\frac{k^{2}}{4e^{2}}\frac{B^{2}}{ |\varphi|^{2}}\frac{1}{w} &=&V
\label{cesb2}
\end{eqnarray}
where 
\begin{eqnarray}
D_{\pm }\varphi &=&\left( D_{1}\pm iD_{2}\right) \varphi \\
|
D_{\mu}\varphi|^{2}&=&-|D_{1}\varphi|^{2}-|D_{2}\varphi|^{2}+e^{2}A_{0}^{2}%
\left\vert \varphi \right\vert ^{2}  \label{X}
\end{eqnarray}
with 
\begin{equation}
A_{0}=\frac{k B}{2 e^2 \left\vert \varphi \right\vert ^{2}\, w}\text{ .}
\label{gl1}
\end{equation}

We now look for vortex solutions of the Eqs.~(\ref{cesb1}) and (\ref{cesb2}%
), i.e., BPS vortices, for which we take the {\textit{Ansatz}} 
\begin{eqnarray}
\varphi (r,\theta )&=& g\left( r\right) \exp \left( in\theta \right)
\label{a1} \\
e\,A^i&=&\epsilon_{i j} \left[ a\left( r\right) -n\right] \, (\widehat{r }^j
\, / r)  \label{a2}
\end{eqnarray}
where $r$ and $\theta$ are the polar coordinates and $n$ is the vortex
winding number. For simplicity, we set $n=1$ from now on.

Inside the core, i.e., near the origin, the boundary conditions are: $%
g(r\rightarrow0)\rightarrow0$, $a(r\rightarrow0)\rightarrow 1$ and $%
A^{\prime}_{0} (r\rightarrow0) \rightarrow 0$, while faraway from it the
vortex fields approach the vacuum, i.e., $g(r\rightarrow\infty)\rightarrow1$%
, $a(r\rightarrow\infty)\rightarrow 0$ and $A^{\prime}_{0}
(r\rightarrow\infty) \rightarrow 0$. This means that the potential has to
have spontaneous symmetry breaking, as usual.

The magnetic field, with magnitude 
\begin{equation}
B=\frac{1}{er}\frac{da}{dr}
\end{equation}%
is parallel to the magnetic momentum, $\mu$, $(\frac{1}{2} \int dr^2 \,
\epsilon_{a b} \, r^a\, j^b)$ and the angular momentum, $L$, $(\int
dr^2\,\epsilon^{a b}\,r_a\,T_{0 b})$. They all vanish near the origin and
faraway from it.

Also note that the magnetic flux, $\phi=2\pi \int dr \, r B(r)$, and the
electric charge, $Q=2\pi \int dr \, r \rho(r)$ are quantized according to 
\begin{equation}
\phi=-\frac{2\pi}{e}\text{ \ \ \ \ \ and \ \ \ } Q=k\phi\text{.}
\end{equation}

Substituting the {\textit{Ansatz}} Eqs.~(\ref{a1}) and (\ref{a2}) into Eq.~(%
\ref{cesb1}) and (\ref{cesb2}) we get the BPS vortex equations given by 
\begin{eqnarray}
\frac{dg}{dr}&=&\frac{g\,a}{r}  \label{cesb0} \\
\frac{k^{2}}{4 e^{2} g^{2}}\frac{B^2}{w}&=&V  \label{cesb40}
\end{eqnarray}
which substituted into Eq.~(\ref{X}) gives 
\begin{equation}
|D_\mu\varphi|^2=-2 \left(\frac{dg}{dr}\right)^{2}+\frac{V}{w} \text{ .}
\label{aX1}
\end{equation}

We note that solutions of the first order equations Eqs.~(\ref{cesb0}) and (%
\ref{cesb40}) do satisfy the second order equations of motion (\ref{maria1})
and (\ref{sf3}).

For further reference we also need to write the Eqs.~(\ref{gl0}) and (\ref%
{sf3}) as 
\begin{eqnarray}
&&\frac{d}{dg}\left[\frac{\sqrt{V/w}}{g}\right]=-\frac{2\,e^2}{k}\,w\,g
\label{eeV} \\
&&\frac{d \left(k A_0\right)}{dr} = 2 \, e\, w\,a\, \frac{g^{2}}{r}
\label{tempgauge}
\end{eqnarray}
which in particular gives that $V$ and $w$ are not independent.

We also need to write the energy density, the spatial component of the
current density, the magnetic and angular momenta which respectively are
given by 
\begin{eqnarray}
\varepsilon&=&2 w\left(\frac{dg}{dr}\right)^{2}+2V  \label{energinaocan} \\
j^{\theta} & =& 2 \, e\, w\,a\, \frac{g^{2}}{r} \\
\mu^z & =& 2 \pi \, e\, \int \, dr\, r \, w\,a \,g^2 \\
L^z & =& 4 \pi \, e\, \int \, dr\, r \, w\,a \,g^2 \, A_0
\end{eqnarray}

We note that the procedure used in this section, to get to the first order
equations, started with the conditions $\mathcal{P}_1=\mathcal{P}_2=0$. This
is motivated by ref. \cite{Vega1}, and it shows explicitly that the choice
of the potential $V(\varphi)$ depends on the choice of $\omega(\varphi)$.
Thus, neither the potential nor the function $w(\left\vert \varphi
\right\vert)$ used to generalize the Chern-Simons model are arbitrary
functions anymore.


\section{Standard self-dual vortices}

In this section we review the vortex solution of the standard Chern-Simons
model that is recovered by setting $w(|\varphi|)\to1$. We use the Eq.~(\ref%
{eeV}) to get 
\begin{equation}
\frac{g}{2}\frac{dV}{dg} = V-\frac{2 e^2 }{k} \, g^3\, \sqrt{V}  \label{evoV}
\end{equation}%
whose solution gives the potential of the standard Chen-Simons model, which
is 
\begin{equation}
V_{S} = \frac{e^4}{k^2}\, g^2 \left(1-g^2\right)^2.  \label{Vcesb}
\end{equation}
Here we have adjusted the integrating constant according to the vortex
boundary conditions.

The first order equations (\ref{cesb0}) and (\ref{cesb40}) then become 
\begin{eqnarray}
\frac{dg}{dr}&=&\frac{a}{r}g  \label{cesb3} \\
e B&=&-\frac{2e^{4}}{k^2}g^2 \left(1-g^2\right)  \label{cesb4}
\end{eqnarray}
which can be integrated numerically from the infinity up to the origin. For
this, we need to write down the asymptotic solutions at the infinity which
are 
\begin{eqnarray}
g(r\rightarrow \infty)&=& 1-C K_0 (mr)  \label{asymp1} \\
P(r\rightarrow \infty)&=& C m r K_1 (mr)  \label{asymp2}
\end{eqnarray}
with $K_i$ the modified Bessel functions, $m={2e^2}/{k}$ and $C$ a constant
which can be adjusted to get the suitable boundary conditions at the origin.

Also note that as 
\begin{equation}
eA_0=\frac{m}{2}\left(g^2-1\right) ,  \label{asymp13}
\end{equation}%
it comes that near the origin $A_{0} (r\rightarrow0) \rightarrow -e/k$ while
faraway from it $A_{0} (r\rightarrow\infty) \rightarrow 0$.

In the figures, we set $m=1$ and plot the Higgs potential, the Higgs and
gauge fields, and the electric and magnetic fields. For that we used the
asymptotic solutions 
\begin{eqnarray}
eB(r\rightarrow \infty)&=&-m^{2} \,C \,K_0 (mr)  \label{asymp3} \\
eE(r\rightarrow \infty)&=&m^{2}\,C \, K_1 (mr)  \label{asymp4} \\
\frac{1}{e}\rho(r\rightarrow \infty)&=&-2 m\,C\,K_0 (mr)  \label{asymp8} \\
\frac{1}{e}j^\theta(r\rightarrow \infty)&=&2 m\,C\, K_1 (mr) \text{.}
\label{asymp9}
\end{eqnarray}
For further reference it is also necessary to write the asymptotic solutions
for the energy density, magnetic and angular momenta given by 
\begin{eqnarray}
\varepsilon(r\rightarrow \infty)&=& 2m^{2} \, C^{2}\,\left[K_0^{2} (mr)
+K_1^{2} (mr) \right]  \label{asymp10} \\
\frac{1}{e}\frac{d \mu^z}{dr}(r\rightarrow \infty)&=&2\pi m C\,r^{2} K_1(mr)
\label{asymp11} \\
\frac{d L^z}{dr}(r\rightarrow \infty)&=&-4\pi m^2 C^2\,r^{2} K_1(mr)\,
K_0(mr)\text{.}  \label{asymp12}
\end{eqnarray}


\section{Generalized self-dual vortices}

In this section we give an example of a vortex solution for the model
introduced in Sec.~II. In order to make a choice for $w$ we first note from
eq. (\ref{eeV}) that if $w$ is not a constant it changes the position of the
zeros of the potential and its maximum amplitude when compared with those
for the standard Chern-Simons potential. We choose $w$ such that the zeros
of $V$ are the same as the zeros of the Higgs potential of the standard
Chern-Simons model. A possible choice for $w$ is $w=3\,(1-g^2)^2$ which,
from Eqs.~(\ref{eeV}) and (\ref{tempgauge}), gives 
\begin{eqnarray}  \label{gpot}
V=\frac{3\,e^4}{k^2}\,g^2\,\left[1-g^2\right]^8 \\
eA_0 = -\frac{e^2}{k}\left[1-g^2\right]^3\text{.}
\end{eqnarray}
Thus, the electric field is given by 
\begin{equation}
e\,E = - \frac{6\,e^2}{k}\,\,\frac{g^2\,a\,\left[1-g^2\right]^2}{r}
\end{equation}
while the first order equations (\ref{cesb0}) and (\ref{cesb40}) become 
\begin{eqnarray}
\frac{dg}{dr}&=&\frac{g\,a}{r} \\
e\,B&=&-\frac{6\,e^4}{k^2}\,g^2\,\left[1-g^2\right]^5\text{.}
\end{eqnarray}

The energy density, the polar component of the current density, the magnetic
and angular momenta are respectively given by 
\begin{eqnarray}
\varepsilon&=& 6\,\left[1-g^2\right]^2\,\left[\left(\frac{dg}{dr}\right)^{2}+%
\frac{e^4}{k^2}\,g^2\,\left[1-g^2\right]^6\right] \\
\ j^{\theta}&=&6\, e\,\frac{g^{2}\,a\,\left[1-g^2\right]^2}{r} \\
\mu^z & =& 6\pi \, e\, \int \, dr\, r \,a \,g^2\,\left[1-g^2\right]^2 \\
L^z & =& -\frac{\pi\,k}{e^2}
\end{eqnarray}

Note that this particular choice for $w$ allows the existence of vortices.
In fact, the vacuum manifold of the Higgs potential is a dot and a circle
which are not simply connected and the energy density is localized (see \cite%
{Vilenkin}). In particular the electric and magnetic field vanish near the
origin and faraway from it. Note that faraway from the origin the vortex
solution approaches the standard vortex solution and therefore the
asymptotic solutions are also given by the Eqs.~(\ref{asymp1}), (\ref{asymp2}%
), and (\ref{asymp3})-(\ref{asymp12}). Also note that there is no divergence
in any physical quantity. All this can be seen from the plots in
Figs.~[1]-[6], where we show and compare the generalized vortex solution
with the one of the standard Chern-Simons model.

\begin{figure}[tbp]
\includegraphics[scale=0.6]{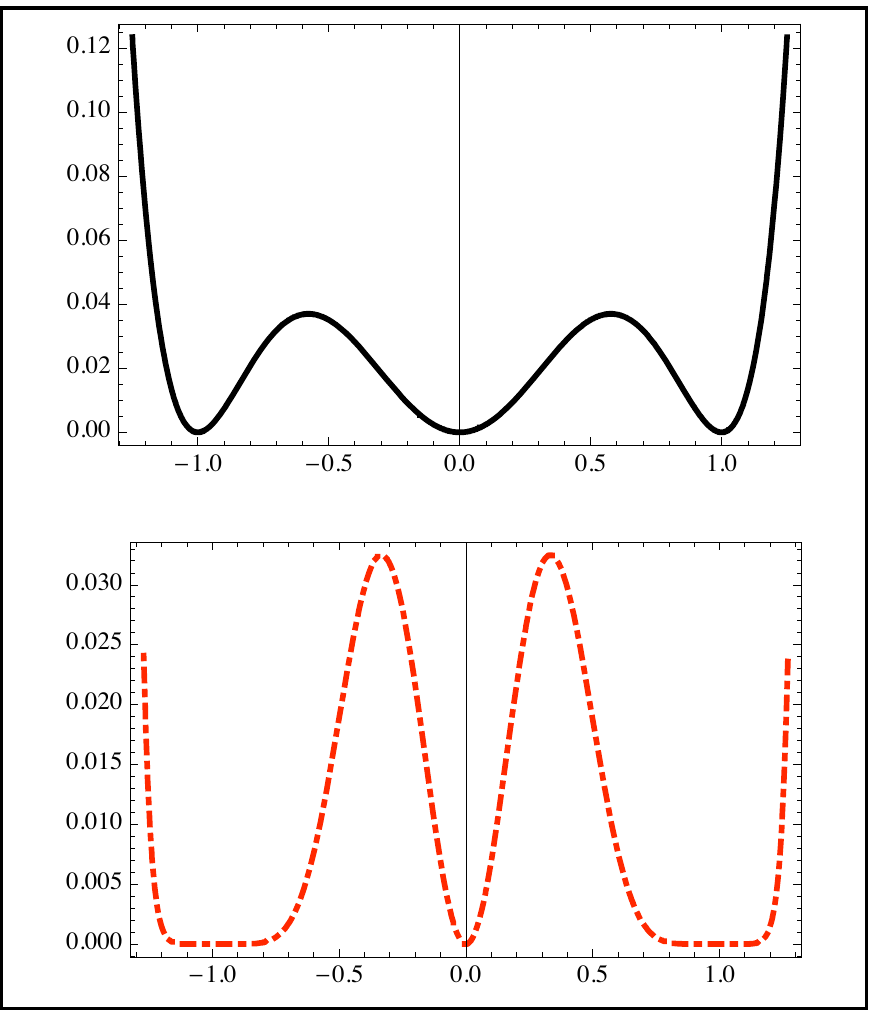}
\caption{The potentials for the standard and generalized models, plotted in
function of the Higgs field, with black/higher and red/lower lines,
respectively.}
\label{Fig1}
\end{figure}

\begin{figure}[tbp]
\includegraphics[scale=0.6]{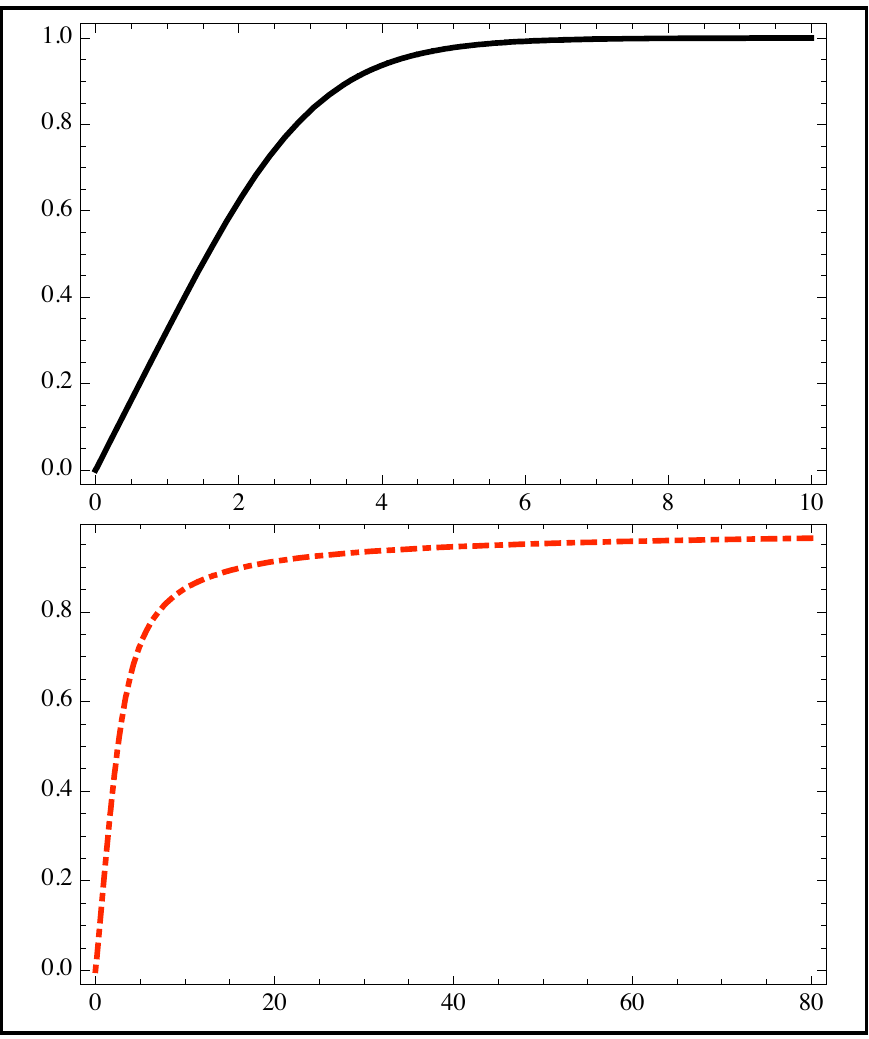}
\caption{The Higgs field as a function of the distance $r$ for the standard
and generalized models. Conventions as in Fig1. Note that in our model the
vacuum is reached for a larger distance than in the standard model -- see
Fig.3 as well}
\label{Fig2}
\end{figure}

\begin{figure}[tbp]
\includegraphics[scale=0.6]{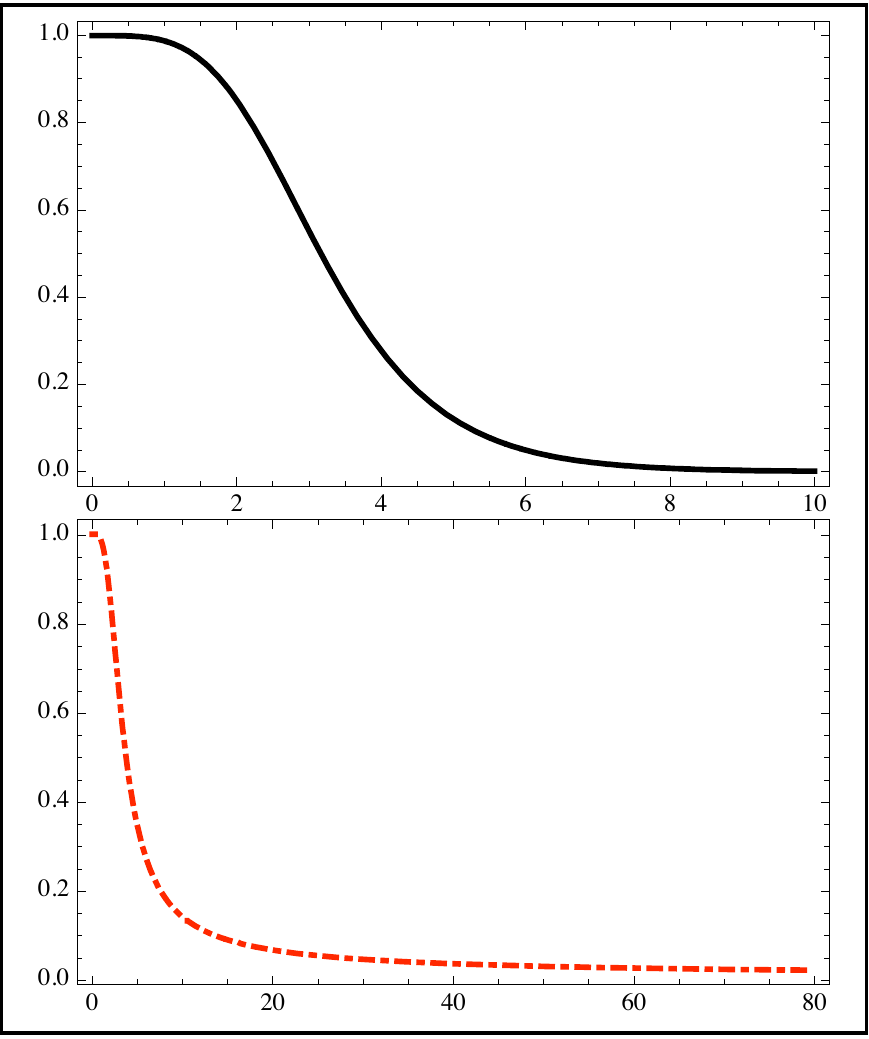}
\caption{The spatial component of the gauge field. Conventions as in Fig.~{%
\protect\ref{Fig2}}.}
\label{Fig3}
\end{figure}

\begin{figure}[tbp]
\includegraphics[scale=0.6]{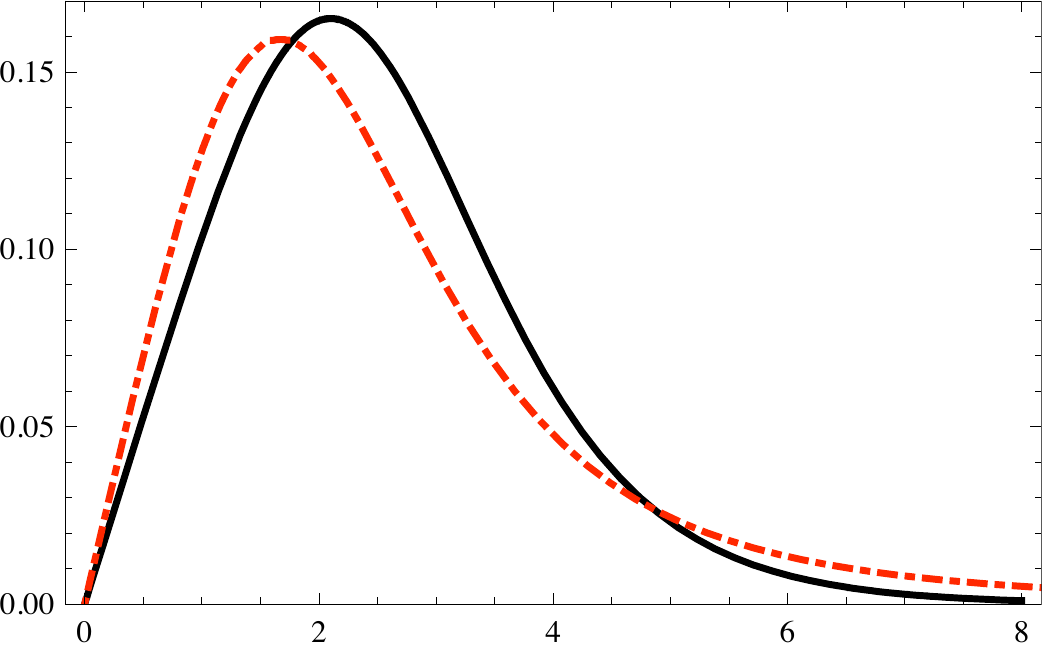}
\caption{The electric field in units of $e$ (black/ red-dotdashed lines for
standard/our model). This is also the Figure for the polar current density $%
j^\protect\theta$ in units of $k$. Note that it is smaller and spread over a
larger distance than in the standard model.}
\label{Fig4}
\end{figure}

\begin{figure}[tbp]
\includegraphics[scale=0.6]{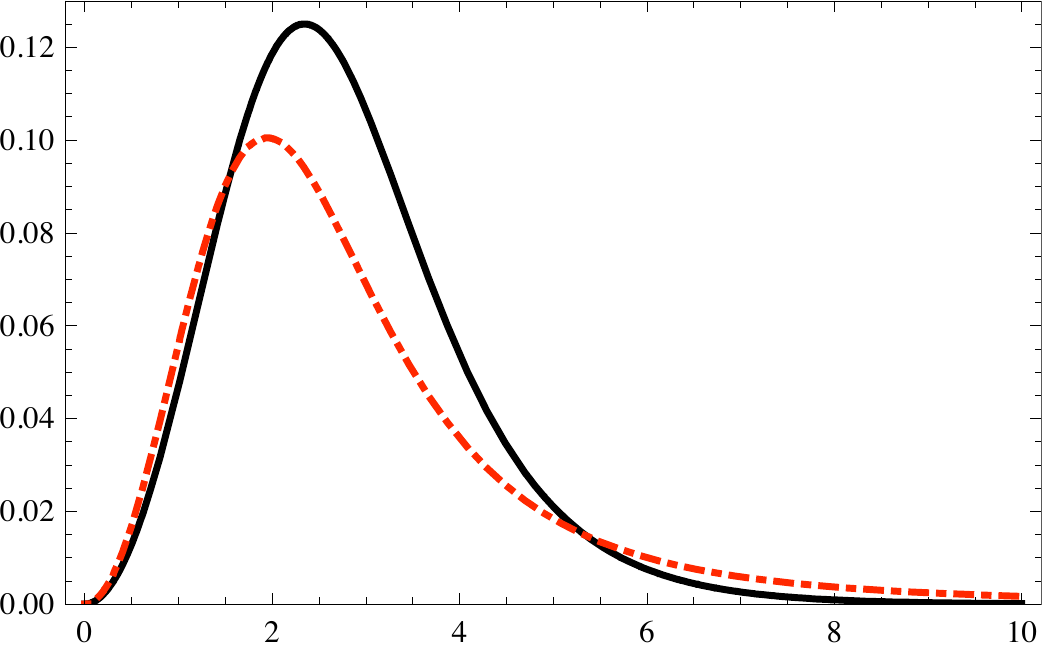}
\caption{The magnetic field in units of $e$. Note that it is smaller and
spread over a larger distance than in the standard model. The same result
applies for the charge density $\protect\rho$. Conventions as in Fig.~{%
\protect\ref{Fig4}}.}
\label{Fig5}
\end{figure}

\begin{figure}[tbp]
\includegraphics[scale=0.6]{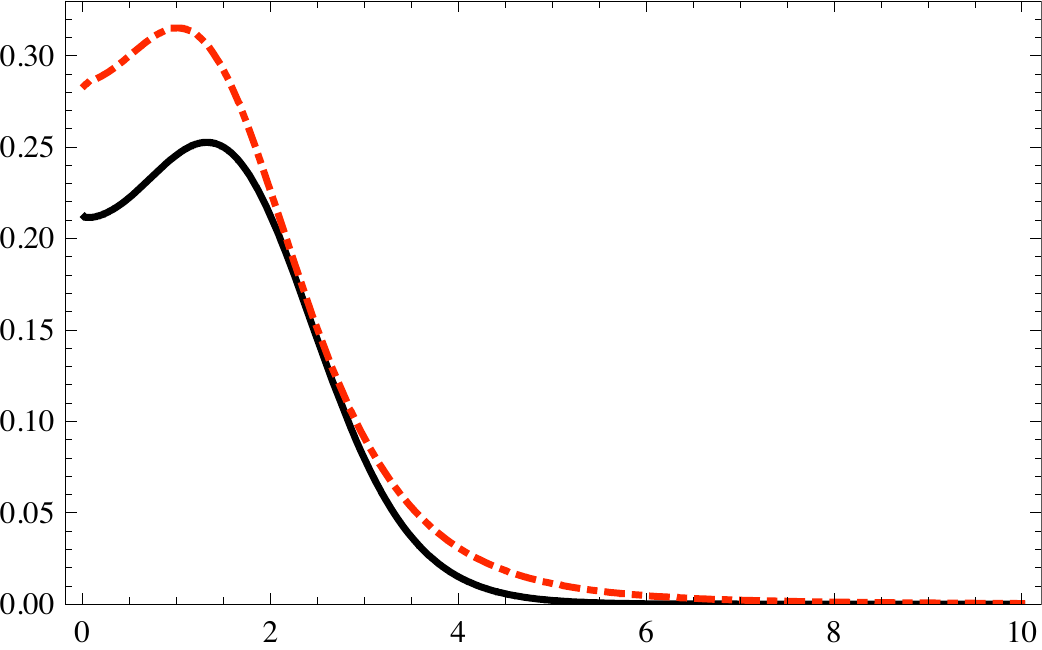}
\caption{The energy density. Conventions as in Fig.~{\protect\ref{Fig1}}.}
\label{Fig6}
\end{figure}


\section{Ending comments}

In this work we have studied the presence of vortices in a generalized
Chern-Simons model. The idea is different from the recent work \cite%
{Babichev1}, where the author has investigated vortices in the Maxwell-Higgs
model, modified to accommodate generalized structure, with the kinetic term
being changed to a function of it. This study was done with the numerical
integration of the equations of motion, and the modification there
introduced has induced another mass scale.

Here, our objective was to generalize the Chern-Simons model in a way such
that we could find first order differential equations. To do this, we have
changed the kinetic term $|D_\mu\varphi|^2$ to $w(|\varphi|)|D_\mu\varphi|^2$%
. We suppose that this modification leads to an effective planar field
theory somehow similar to the standard Chern-Simons model. Despite the
modification in the kinematic scalar field term, we could write a first
order framework and find vortices which are qualitatively similar to the
vortices of the standard Chern-Simons model. However, we could identify
several properties of the BPS vortices which are quantitatively different
from the standard vortices, since the solutions can be thicker then the
standard solutions. These differences are shown in all the figures, where we
depict distinct features of the vortices in both the generalized and
standard Chern-Simons models.

When compared to the model investigated in \cite{Babichev1}, an important
distinction which appears in our work is that the modification we have
included does not introduce another mass scale in the system. To see this
clearly, let us write the potential in Eq.~(\ref{gpot}) in terms of
dimensional quantities. It writes 
\begin{equation}
V(|\varphi|)=\frac{3\,e^4}{k^2v^6}|\varphi|^2(v^2-|\varphi|^2)^8
\end{equation}
where $v$ is the symmetry breaking parameter of the model. In this case, the
mass scale $M$ which we had to include at the end of Sec.~II can be seen as $%
M=v^2$, so we do not need an extra mass scale, which had to be included in 
\cite{Babichev1}.

We are now examining how to obtain first order equations in a more general
model, modifying the kinetic scalar field term but including both the
Maxwell and the Chern-Simons terms. Also, we are studying the presence of
vortices in a Maxwell-Higgs model with the k-field modification similar to
the case investigated in \cite{Babichev1}. Preliminary results indicate the
presence of compact vortices, e.g., of vortices with the scalar and gauge
fields getting to their vacuum values at finite distances from the origin.

Before ending the work, let us study the Bogomol'nyi decomposition of the
energy of the static solutions of the first order equations found above. To
make the calculation explicit, we rewrite the energy density (\ref{csed2})
in the form 
\begin{eqnarray}
\varepsilon  &=&w\left\vert \left( D_{1}\pm iD_{2}\right) \varphi
\right\vert ^{2}+\left( \frac{kB}{2eg\sqrt{w}}%
\mp \sqrt{V}\right) ^{2}  \notag \\
&&\;\;\;\pm ewg^{2}B\pm \frac{kB}{eg}\sqrt{%
\frac{V}{w}}\pm \frac{w}{r}\frac{d}{dr}\left( g^{2}a\right) \text{ .}
\end{eqnarray}%
This result can be used to recover the standard case of the Chern-Simons
model. We make $w\left( g\right) =1$ to get to 
\begin{eqnarray}
\varepsilon  &=&\left\vert \left( D_{1}\pm iD_{2}\right) \varphi \right\vert
^{2}+\left( \frac{kB}{2eg}\mp \frac{e^{2}}{k}g\left( g ^{2}-1\right) \right)
^{2}  \notag \\
&&\;\;\;\pm eB\pm \frac{1}{r}\frac{d}{dr}\left( g^{2}a\right) \text{ ,}
\end{eqnarray}%
with the standard potential 
\begin{equation}
V=\frac{e^{4}}{k^{2}}g^{2}\left( 1-g^{2}\right) ^{2}\text{,}
\end{equation}%
which leads to the first order equations obtained in Sec.~III.

On the other hand, if we take $w\left( g\right) =3\,(1-g^{2})^{2}$ we get 
\begin{eqnarray}
\varepsilon  &=&3\left( 1-g^{2}\right) ^{2}\left\vert \left( D_{1}\pm
iD_{2}\right) \varphi \right\vert ^{2}\mp \frac{1}{r}\frac{d}{dr}\left(
a\left( 1-g^{2}\right) ^{3}\right)   \notag \\
&&+\left( \frac{kB}{2eg\sqrt{3}\left( 1-g^{2}\right) }\mp \frac{\sqrt{3}e^{2}%
}{k}g\left( 1-g^{2}\right) ^{4}\right) ^{2}\text{ ,}
\end{eqnarray}%
where we used 
\begin{equation}
V=\frac{3e^{4}}{k^{2}}g^{2}\left( 1-g^{2}\right) ^{8}\text{ .}
\end{equation}
We point out that an integration of $r^{-1}d_{r}\left( a\left(
1-g^{2}\right) ^{3}\right)$ over all planar space can be
identified with an integration of $-B(r)$ the same space.
This integration process gives the magnetic flux $\phi$ which is
topologically invariant. In this way, this result leads to the first order
equations used above, so the corresponding solutions are in fact BPS states,
with the energy bound being $E_{B}=e\,|\phi |$ where $\phi $ represents the
flux of the magnetic field in the plane. We note that the energy bound in
the generalized model is the same of the energy bound of the standard
Chern-Simons model.

\section{Acknowledgments}

We thank CAPES and CNPq, Brazil and Christoph Adam and Filipe Correia for interesting comments.
C dos Santos is partially financed by 
SFRH/BSAB/925/2009, FCT Grant No. CERN/FP/109306/2009.
and would like to thank the Departamento de Fisica USC for all their hospitality
while doing this work.


\end{document}